# Low frequency communication based on Rydberg-atom receiver

YIPENG XIE [1,2], MINGWEI LEI[1], MENG SHI[1,*]

[1]*Key Laboratory of Space Utilization, Technology and Engineering Center for Space Utilization, Chinese Academy of Sciences, Beijing 100094, China*

[2]*University of Chinese Academy of Sciences, Beijing 100049, China;*

*\* shimeng@csu.ac.cn*

## Abstract

Low frequency communication has a wide range of applications in the fields of satellite detection, underground mining, disaster relief. Rydberg atom sensor has rapidly developed in recent years, capitalizing on its calibration-free SI-traceability, large polarizabilities and transition dipole moments. A Rydberg atom sensor is capable of sensitively detecting electric field signals from DC to THz. In this work, we demonstrate low frequency communication using Rydberg atoms in a vapor cell with two parallel electrode plates inside. Three modulations, BPSK, OOK, and 2FSK, are used for the communication by Rydberg atom receiver near 100kHz. We have measured the SNR of the modulated low frequency signal received by Rydberg atoms at various emission voltages. Meanwhile, we have demonstrated IQ constellation diagram, EVM and eye diagram of the demodulated signal at different symbol rate. The EVM is measured to be 8.8% when the symbol rate is 2Kbps, 9.4% when the symbol rate is 4Kbps, and 13.7% when the symbol rate is 8Kbps. The high-fidelity digital color image transmission resulted in a peak signal-to-noise ratio of 70dB. This study proves that Rydberg-atom receiver can finely work in low frequency communication.

## 1. Introduction

Frequency of electromagnetic wave in low frequency communication is less than 300 KHz[1]. Low frequency signals have many advantages in propagation, such as low loss, high bypass capability, long range, and low absorption loss in seawater and soil[2]. However, in the realm of low frequency communication, the acquisition of high-sensitivity reception is contingent upon the deployment of extensive antenna arrays. The spatial constraints at the receiver's location may render the implementation of such large-scale antenna systems impracticable. Consequently, this necessitates the exploration of alternative strategies to enhance reception sensitivity without the requirement for disproportionately large antenna structures. Rydberg-atom-based receiver sized in centimeters, decoupling from the electric field frequency[3], holds significant promise for advancements in low frequency communication.

The Rydberg atoms exist in a state of high energy, characterized by their calibration-free SI-traceability, large polarizabilities and transition dipole moments[4]. A Rydberg atom sensor is capable of sensitively detecting electric field signals from DC to THz[5]. Utilizing the EIT-AT effect or Stark shift effect[6, 7] enables the measurement of factors such as the amplitude, phase, and polarization of the electric field[8, 9]. Researchers have dedicated their efforts to creating Rydberg-atom-based receivers covering from 300 MHz to 300 GHz [10-16]. Additionally, video and image signal has been received by the Rydberg-atom-based receivers successfully[17, 18]. Current communication technologies based on Rydberg atoms predominantly operate in the high-frequency band, with relatively few endeavors exploring the low-frequency spectrum. The shielding effect of the alkali-metal-atom container on low-frequency electric fields [19] makes it difficult for low frequency signals below MHz to be detected. Such obstacles impede the development of low frequency communication receivers utilizing Rydberg atoms. There are two ways to overcome this disadvantage. One is making the vapor cell by sapphire[20, 21], which has a much higher internal surface resistance than glass. However, this kind of method incurs a significantly high cost. The other one is placing two parallel electrode plates inside the vapor cell and the low frequency signal can be lead in the cell directly by the two plates[22]. This methodology, while cost-effective, may compromise the hermetic seal of the vapor cell.

We have addressed the hermetic seal issue associated with the second method by employing a specialized fabrication technique. In this paper, we demonstrate low frequency communication using Rydberg atoms in a vapor cell containing two parallel electrode plates. We first study the spectrum of the signal received by Rydberg atoms under the three modulation modes such as OOK, PSK, FSK. After that we demonstrate the signal-to-noise ratio of the received signal as a function of various emission voltage. Subsequently, we test different parameters at different symbol rate to evaluate the quality of Rydberg-atom-based low frequency communication, such as IQ constellation diagram, Error Vector Magnitude (EVM) and eye diagram. We also show high-fidelity digital image transmission based on the low frequency communication based on Rydberg atoms, yielding a peak signal noise ratio (PSNR) of 70dB for image transmission fidelity. This study demonstrates the great potential of Rydberg atoms in the field of low frequency communication.

## 2. Experiment setup

We perform the experiment in a cylindrical cesium room-temperature vapor cell with $100\ mm$ long and $40\ mm$ diameter. The experimental setup and the relevant three-level Rydberg-EIT diagram are illustrated in Fig. 1 (a) and (b). An 852 $nm$ laser with diameter of $600um$ is split into two beams, one is the probe beam ($\lambda_p$) with a power of $50uW$, and the other one is an identical reference beam. Both are propagating in parallel through the cell. A coupling laser ($\lambda_c$=509 $nm$) with power of $15mW$ and diameter of 1mm counter-propagates and overlaps with the probe laser beam. The probe laser drives the transition of $|6S_{1/2}, F=4>\rightarrow|6P_{3/2}, F'=4,5>$ and the coupling laser is scanned through the Rydberg transition of $|63D_{5/2}\rangle$ from $|6P_{3/2}3/2, F'=4,5\rangle$, thus establishing Rydberg-EIT spectroscopy to enhance the probe transmission at two-photon resonant condition. Besides, the frequency of the probe laser and coupling laser are both locked to an ultra-stable cavity (PDH) to reduce the laser frequency noise. The transmission EIT signal is detected with a differential photodetector (Thorlabs PDA210A/M) as a transmission difference with the reference beam, eliminating the intensity noise of the probe laser. The EIT signal, after being displayed and captured by an oscilloscope (SDS5000X, SIGLENT), are stored for subsequent demodulation and data recovery using a MATLAB program executed on a personal computer. A pair of coppery electrode plates (size of 80 $mm$ × 26 $mm$ × 1 $mm$) is parallel integrated into the cesium cell with a spacing of 18 $mm$. Each the coppery plate has a copper pillar passing through the vapor cell. One of the copper pillars connects to a whip antenna to couple low frequency signal to vapor cell, and the other one connects to the ground.

The resistor with a value of 10$Ω$ used to transmit the low frequency signal is connected to the signal generator (DG2000, RIGOL) through the OPA544 amplifier. The signal generator employed can provide a variety of low frequency modulated signals. The output voltage of the generator spans a range from 0 to 20 volts. The role of the OPA544 operational amplifier is to augment the power radiated by the resistance into the surrounding space by increasing the output current. Therefore, the emission power is observed to increase linearly with the augmentation of the voltage. Although the radiative efficiency of the low-frequency signals in this research is not very high, the signals emitted in this manner are still capable of being detected by Rydberg-atom receiver within a defined distance and within a specific voltage threshold.

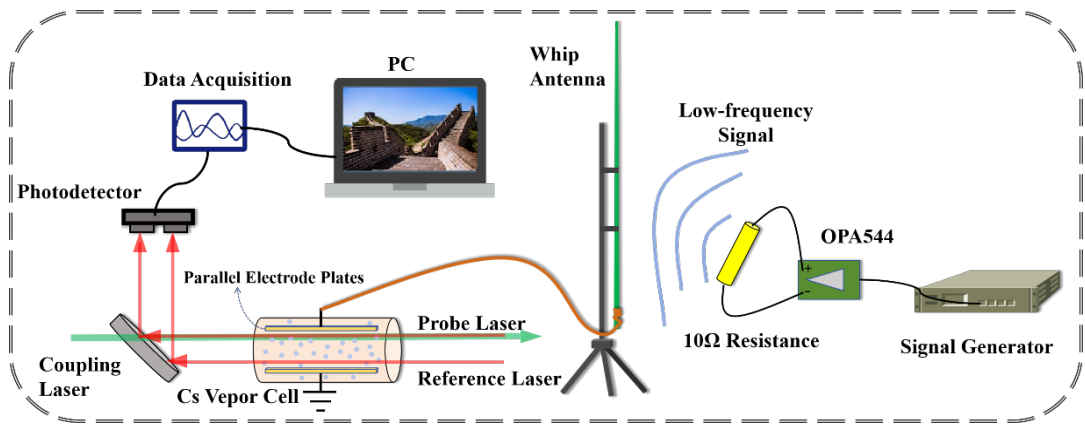

**Fig. 1.** Schematic of the experimental setup. An 852 nm probe laser and an identical reference beam are both propagating in parallel through the vapor cell, where a pair of coppery electrode plates is parallel integrated into the cell with a spacing of 18 mm, while a 509 nm Rydberg laser counter-propagate and overlap with the probe laser, but not the reference beam. The transmission of the probe and reference beam are incident into a differential photodetector as a transmission difference.

## 3. Results and analysis

In this study, we employed three distinct modulation techniques, specifically 2FSK (Two-Frequency Shift Keying), OOK (On-Off Keying), and BPSK (Binary Phase Shift Keying). The typical waveforms of the three modulations are shown in Fig2. (a). In 2FSK, two distinct frequencies are used to represent the two possible states of a digital signal, typically denoted as '1' and '0'. The transition between these frequencies is instantaneous, and the frequency remains constant for the duration of each bit interval, thus encoding the binary information into the frequency domain. OOK is characterized by the transmission of a continuous wave (CW) signal during the '1' state and the absence of the signal during the '0' state. This method is particularly susceptible to noise and requires a relatively high signal-to-noise ratio (SNR) to achieve reliable detection.

In BPSK, two phases are used, typically 180 degrees apart, to symbolize the two binary digits. The phase transition is binary, and the phase state remains constant for the duration of each bit interval, encoding the digital information into the phase domain.

We use the PRBS (Pseudo-Random Binary Sequence) signal, as endorsed by the International Telecommunication Union (ITU), for low frequency communication test. Shown as Fig2. (b), PRBS are generated through a deterministic linear feedback shift register (LFSR) process, which emulates the properties of randomness. The generation of PRBS involves a sequence of binary states produced by the feedback of the output of the shift register through a series of taps, which are predefined by a characteristic polynomial. This polynomial dictates the length of the shift register and the period of the resulting sequence, ensuring a high degree of predictability and repeatability, while still exhibiting the statistical properties akin to random sequences. The generation polynomial in this work is $1 + x^6 + x^7$.

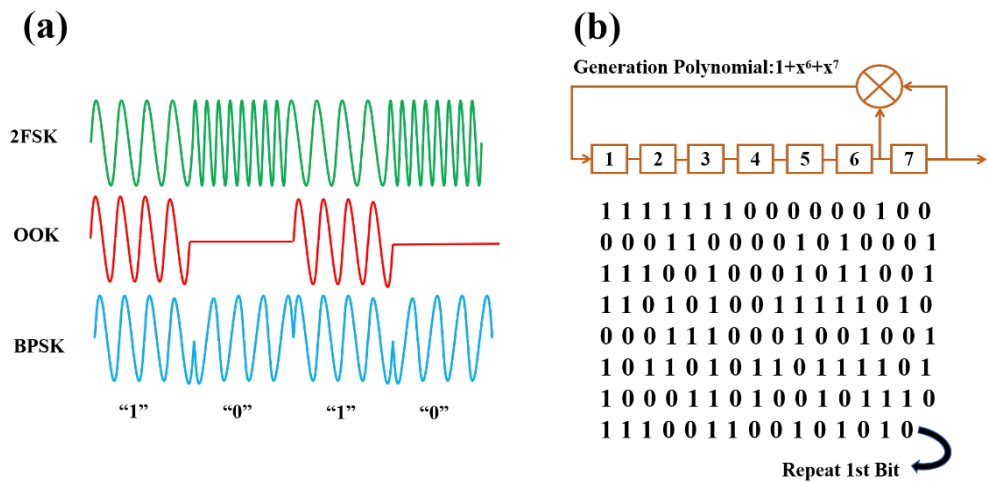

**Fig.2.** (a)Waveforms of 2FSK, OOK, BPSK signal. 2FSK involves modulating the carrier signal's frequency to represent two binary states, with two distinct frequencies corresponding to '1' and '0'. OOK is an amplitude modulation technique where the presence of the carrier signal indicates a binary '1' and its absence indicates a binary '0'. BPSK is characterized by a carrier signal whose phase is modulated to represent two distinct binary states, typically 0 and 1. (b) The generation process of PRBS. The generation polynomial in this work is 1+x⁶+x⁷.The total bits in one period is 127.

We need to obtain the spectrum of the signal received by Rydberg atoms under the three modulation modes, 2FSK, OOK, and BPSK, to demodulate the signal precisely as well as to calculate communication parameters such as SNR (signal-to-noise ratio) and BER (bit error rate). We set a data rate($f_b$) of 2Kbps, a communication distance of 15cm, and a emission voltage of 20v in the test. The spectrums are shown in Fig. 3. In the experiment, the carrier frequency $f_c$ of OOK and BPSK are 100kHz, and the two carrier frequency $f_1$ and $f_2$ of 2FSK are 90kHz and 110kHz. From the spectrum it can be concluded that the OOK signal bandwidth is between 98kHz and 102kHz ($f_c \pm f_b$), the BPSK signal bandwidth lies between 98kHz and 102kHz ($f_c \pm f_b$), and the 2FSK signal bandwidth lies between 88kHz and 112kHz($|f_2 - f_1| + 2f_b$). It is evident from the analysis that the bandwidth of the filters used in the demodulation process should be substantially similar when demodulating OOK signal as compared to BPSK signal.

Among the modulation schemes considered, 2FSK exhibits the greatest bandwidth occupancy at equivalent data rates due to its inherent frequency deviation properties, which necessitate a broader frequency spectrum to distinguish between the two frequency components encoding the binary information.

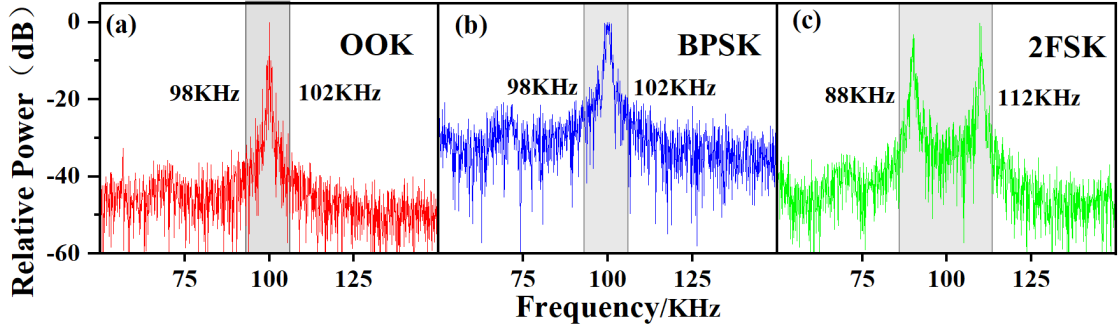

**Fig. 3.** Spectrum for different modulation modes. (a) the OOK signal bandwidth is between 98kHz and 102kHz ($f_c \pm f_b$); (b) the BPSK signal bandwidth lies between 98kHz and 102kHz ($f_c \pm f_b$); (c) the 2FSK signal bandwidth lies between 88kHz and 112kHz ($|f_2 - f_1| + 2f_b$).

In practical communication application, to realize the reliable transmission and successful demodulation, the signals received must reach a certain strength and have a high SNR. Therefore, it is necessary to observe the demodulated signal waveforms and SNR changes under different emission voltages to ensure reliable data transmission and reception. Fig. 4 shows the waveforms of the demodulated PRBS signal received by the Rydberg atom under three modulation modes when the emission voltage is adjusted to 8V, 4V, and 1V, respectively. At an emission voltage of 8V, the demodulation of signals modulated via BPSK, OOK, and 2FSK schemes can accurately reflect the characteristics of the original signals, suggesting an adequate signal-to-noise ratio for effective demodulation. Upon reduction of the emission voltage to 4V, the demodulated signals of BPSK and OOK manifest significant noise interference, whereas the 2FSK signal demonstrates enhanced noise resistance. At an emission voltage of 1V, demodulation of all signal types becomes infeasible, indicating that the noise level has surpassed the threshold necessary for reliable signal recovery

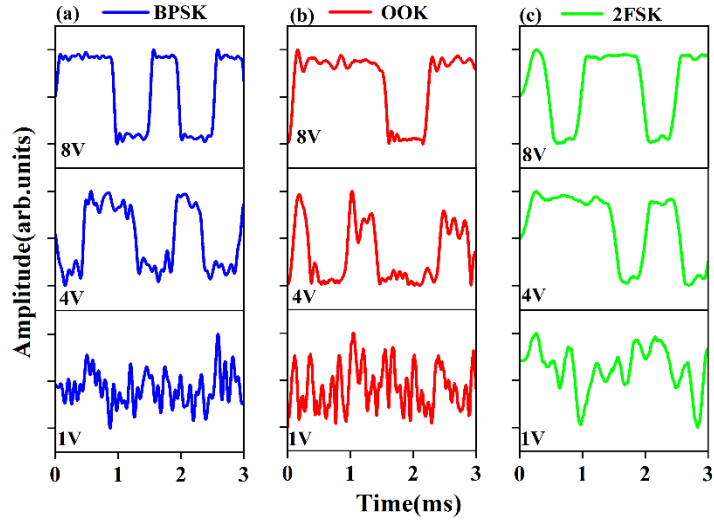

**Fig. 4** Demodulated signal waveforms of BPSK, OOK, and 2FSK at emission voltages of 8V, 4V, and 1V. (a) BPSK; (b) OOK; (c) 2FSK

In the realm of low-frequency communication, the application of nonlinear noise processing techniques is imperative. Moreover, the presence of nonlinear amplification circuits within low-frequency signal transmission systems necessitates the selection of modulation schemes that maintain a constant envelope to mitigate the impact of nonlinearity on the signal. Additionally, given the high-power output of the transmitter, it is crucial to prevent potential high-voltage transient issues between the transmitter and the antenna, thus the transmitted signal should exhibit continuous phase characteristics. Furthermore, as the signal traverses the marine channel, it experiences significant distortion in both amplitude and phase. In this study, three modulation methods, OOK, BPSK and 2FSK, were considered. The OOK does not maintain a constant signal envelope during transmission, and BPSK lacks continuous phase during the transmission process. Consequently, FSK has emerged as a prevalent modulation technique in low-frequency communication.

Shown as Fig. 5, the SNR of the three different signals received is compared as a function of the emission voltage. When the emission voltage is relatively high (ranging from 6 to 10 volts), the SNR of the three types of received signals does not exhibit significant variation with the decrease in emission voltage. Moreover, the SNR of the three types of received signals is nearly identical under the same emission voltage, maintaining a level around 22 dB. As the emission voltage further decreases below 6 volts, the SNR of all three types of received signals experiences a noticeable decline. However, under the same emission voltage, the SNR of 2FSK consistently surpasses that of the other two methods, demonstrating its superiority in low-frequency communication When the SNR falls below the threshold of 4 dB, the demodulation of all three types of received signals becomes unsuccessful.

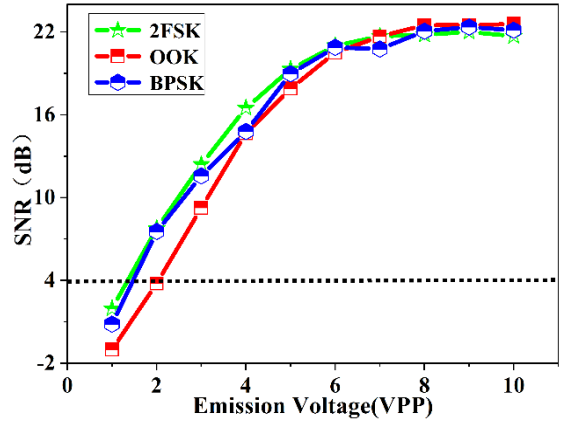

**Fig. 5.** the SNR of the three different signals is compared as a function of the emission voltage. The SNR of the FSK signal received by the atom is higher than that of OOK and BPSK. When the SNR falls below the threshold of 4 dB, the demodulation of all three types of received signals becomes unsuccessful.

In communications, the IQ constellation diagram is a graphical tool used to represent the distribution of phase and amplitude states of a modulated signal. EVM is an error vector of the received phase/amplitude state compared to the ideal state and is an assessment of the received modulation quality. The eye diagram is used to analyze the integrity analysis of communication signals, showing the time delay and amplitude variation of the signal during communication transmission. When the opening of the eye diagram is larger, it indicates a higher BER. From (a), (b), and (c) in Fig. 6, when the symbol rate increases, the signal point density on the BPSK constellation diagram becomes more uneven and more points deviate from the ideal position, which indicates that the SNR of the received signal as well as the BER increases. And the opening of the eye diagram gradually increases as the sign rate rises, as shown in Fig. 6(d)(e)(f). The EVM is measured to be 8.8% when the symbol rate is 2Kbps, 9.4% when the symbol rate is 4Kbps, and 13.7% when the symbol rate is 8Kbps.

Shown as Fig. 7, the EVM of the BPSK signal received is compared as a function of the symbol rate. At high symbol rates, ranging from 16 to 20 Kbps, the EVM exhibits a rapid decline as the symbol rate decreases. Upon further reduction of the symbol rate below the 16 Kbps mark, the rate of EVM decrease slows down, eventually stabilizing in the vicinity of 8.8%. This trend indicates that as the symbol rate increases, the communication quality of low-frequency communication systems employing BPSK as the modulation scheme deteriorates.

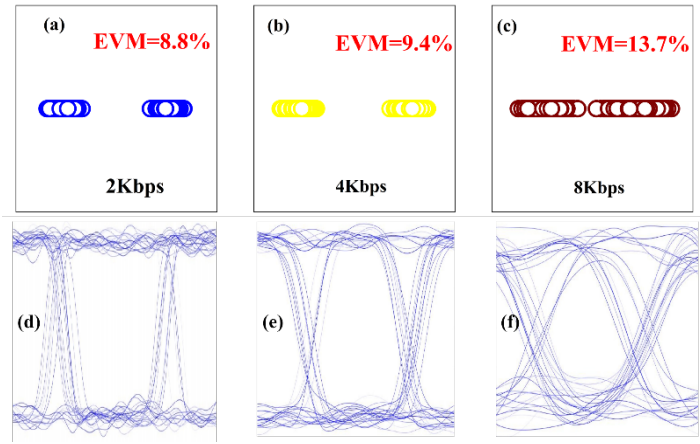

**Fig. 6.** IQ constellation diagrams and eye diagrams of the Rydberg atom reception at different symbol rates. (a) The EVM is measured to be 8.8% when the symbol rate is 2Kbps; (b) The EVM is measured to be 9.4% when the symbol rate is 4Kbps; (c) The EVM is measured to be 13.7% when the symbol rate is 8Kbps; (d) Eye diagrams for symbol rate of 2Kbps; (e) Eye diagrams for symbol rate of 4Kbps; (f) Eye diagrams for symbol rate of 8Kbps.

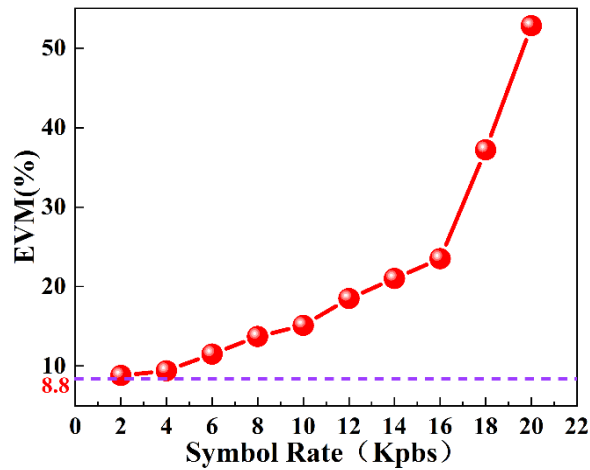

**Fig. 7.** Variation of EVM with symbol rate when BPSK signal is received by Rydberg atoms. As the symbol rate of the BPSK signal gradually increases, the EVM also increases, representing poorer communication quality.

One of the main characteristics of low frequency communication is the low data rate due to the influence of the communication band and environmental factors. Hence, it is usually not used for audio and video transmission. However, in a controlled laboratory setting, we successfully received black-and-white image and color image modulated using 2FSK technique by employing Rydberg atom at a data rate of 10 Kbps. The color image transmission process, depicted in Fig. 8, is principally divided into 6 components: decomposing the original image into its constituent R, G, and B color spaces, mapping each pixel's intensity value to an 8-bit sequence, encoding these sequences into baseband signals with a series of high and low voltage levels, modulating these baseband signals onto a carrier using 2FSK, demodulating the received signals to retrieve the original 8-bit sequences, and ultimately reconstructing the pixel data to recover the final color image.

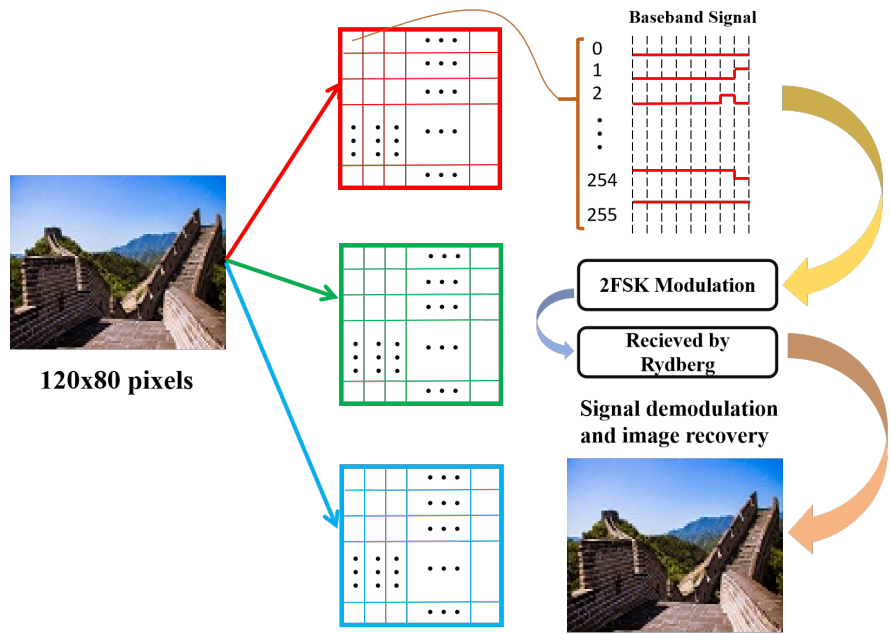

**Fig.8.** The process of color image transmission

The image width is 120 pixels, and the height is 80 pixels. The reception is shown in Fig. 9. From the reception results, both black-and-white and color images can be recovered after received by the Rydberg Atom. From Figure 9, it can be observed that black and white images are fully recoverable, whereas the restored color images exhibit a degree of blurriness. In order to evaluate the quality and reliability of color image transmission, we mainly analyze the fidelity of the image from the peak signal noise ratio (PSNR). The formula for the PSNR is expressed as follows:

$$PSNR = 10 \times \log_{10}(\frac{MAX^2}{MSE}) = 20\log_{10}(\frac{MAX}{\sqrt{MSE}})$$

Where MAX is the maximum pixel value in the image, MSE is a mean square error. Usually, the PSNR is considered excellent if it is higher than 40dB. The calculated PSNR of the color image received by the Rydberg atom is 70 dB.

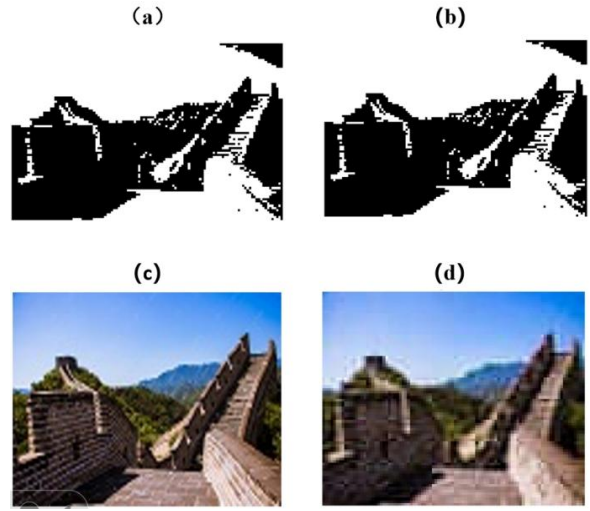

**Fig. 9.** Reception of black-and-white and color images. (a) Original black and white image; (b) Received black and white image; (c) Original color image; (d) Received color image.

## 4. Conclusion

In summary, the presented research successfully integrates Rydberg-atom technology into the domain of low-frequency communication, showcasing its applicability and superiority. The experiments conducted within this study have demonstrated that Rydberg atoms in a vapor cell are capable of effectively receiving signals modulated via 2FSK, OOK, and BPSK, with 2FSK exhibiting slightly better performance in terms of noise resistance and SNR stability. This finding is particularly noteworthy given the challenges associated with low-frequency signal reception. The high-fidelity transmission of digital images, culminating in a peak PSNR of 70 dB, underscores the practical viability of Rydberg-atom-based systems for low-frequency communication. The results indicate that Rydberg-atom receivers are not only sensitive but also robust against environmental interference, a critical attribute for applications such as satellite telemetry, underground communication, and emergency response systems.

**Funding.** This work was supported by the National Natural Science Foundation of China (U2341211).